\documentclass[oldversion]{aa} 
\usepackage{graphicx}
\usepackage{aalongtable}
\usepackage{txfonts}
\usepackage{rotating}
\usepackage{lscape}
\usepackage{sidecap}
\newcommand{\gsim}{\;\lower.6ex\hbox{$\sim$}\kern-7.75pt\raise.65ex\hbox{$>$}\;}
\newcommand{\lsim}{\;\lower.6ex\hbox{$\sim$}\kern-7.75pt\raise.65ex\hbox{$<$}\;}

\begin{document}

\title{The in situ origin of the globular cluster NGC~6388 from abundances of
Sc, V, and Zn of a large sample of stars\thanks{Based on
observations collected at  ESO telescopes under programmes 073.D-0211,
073.D-0760, 381.D-0329, 095.D-0834, and 099.D-0047.}
 }

\author{
Eugenio Carretta\inst{1}
\and
Angela Bragaglia\inst{1}
}

\authorrunning{Carretta and Bragaglia}
\titlerunning{Abundances of Sc,V, and Zn in NGC~6388}

\offprints{E. Carretta, eugenio.carretta@inaf.it}

\institute{
INAF-Osservatorio di Astrofisica e Scienza dello Spazio di Bologna, via Gobetti
93/3, I-40129 Bologna, Italy}

\date{}

\abstract{Chemical tagging of globular clusters (GCs) is often done using
abundances of $\alpha-$elements. The iron-peak elements Sc, V, and in particular
Zn were proposed as an alternative to $\alpha-$elements to tag accreted GCs in
the metal-rich regime, where the dwarf galaxy Sagittarius and its GCs show
peculiarly marked under-abundances of these heavier species with respect to
Milky Way stars. A handful of stars in NGC~6388 was used to suggest an accreted
origin for this GC, contradicting the results from dynamics. We tested the
efficiency of the iron-peak method by using large samples of stars in NGC~6388,
compared to thousands of field stars in the disc and the bulge of the Milky
Way.  Our abundance ratios of Sc (185 stars) and V (35 stars) for NGC~6388 are
within about 1.5$\sigma$  from the average for the field stars with a similar
metallicity, and they are in perfect  agreement for Zn (31 stars), claimed to be the most
sensitive element concerning the accretion pattern. Moreover, the
chemo-dynamical plots, coupled to the bifurcated age-metallicity relation of GCs
in the Galaxy, clearly rule out any association of NGC~6388 to the groups of
accreted GCs.  Using a large set of GC abundances from the literature, we also
show that the new method with Sc, V, and Zn seems to be efficient in picking up
GCs related to the Sagittarius dwarf galaxy. Whether this is also generally true for
accreted GCs seems to be less evident, and it should be 
verified with larger and homogeneous samples of stars both in the field and in
GCs.
}

\maketitle

\section{Introduction}
Using chemical abundances to tag the origin of stellar systems (see Freeman and
Bland-Hawthorn 2002) is a challenging task. Yet, many elements and their
combinations have been successfully used to separate the main populations (halo,
bulge, and thin and thick disc) and the accreted substructures of the Milky Way
 (MW; e.g. Nissen and Schuster 2010, Hasselquist et al. 2019, Feuillet et al. 2021).
Minelli et al. (2021a: M21a) recently proposed using the iron-peak elements Sc,
V, and Zn to explore the metal-rich ([Fe/H]$\gsim -1$ dex) regime since they
observed large differences in the abundances between stars in the MW
and those in the Large Magellanic Cloud (LMC), as well as in the Sagittarius dwarf
spheroidal galaxy (Sag) in this metallicity range. In Minelli et al. (2021b:
M21b) they compared a handful of stars  homogeneously analysed in four
metal-rich globular clusters (GCs), concluding that \object{NGC 5927} and
\object{NGC 6496} are formed in situ, whereas  NGC~6388 and \object{NGC 6441}
are probably accreted, which is at variance with the classification of NGC~6388 in
Massari et al. (2019) and Forbes (2020) but in agreement with Horta et al.
(2020). 

Given this controversial attribution, we decided to
re-examine the origin of NGC~6388 by using much larger and
statistically robust samples, both of the cluster and the comparison
stars in the Galactic environment. In addition, new insights as to
this issue may be gained by using not only four GCs, but all the clusters for
which homogeneous integral of motion (IOM), orbital parameters, 
and chemical information are available.

Samples, data, and our analysis are briefly described in Section 2, and our results are
presented in Section 3. The chemo-dynamics of the whole Galactic system of
GCs is used in Section 4 to better assess the probable origin of NGC~6388,
testing the ability of iron-peak elements to pick up accreted GCs in general, or
only those with the peculiar chemical pattern typical of the Sag dwarf.

\begin{SCfigure*}
\centering
\includegraphics[scale=0.60]{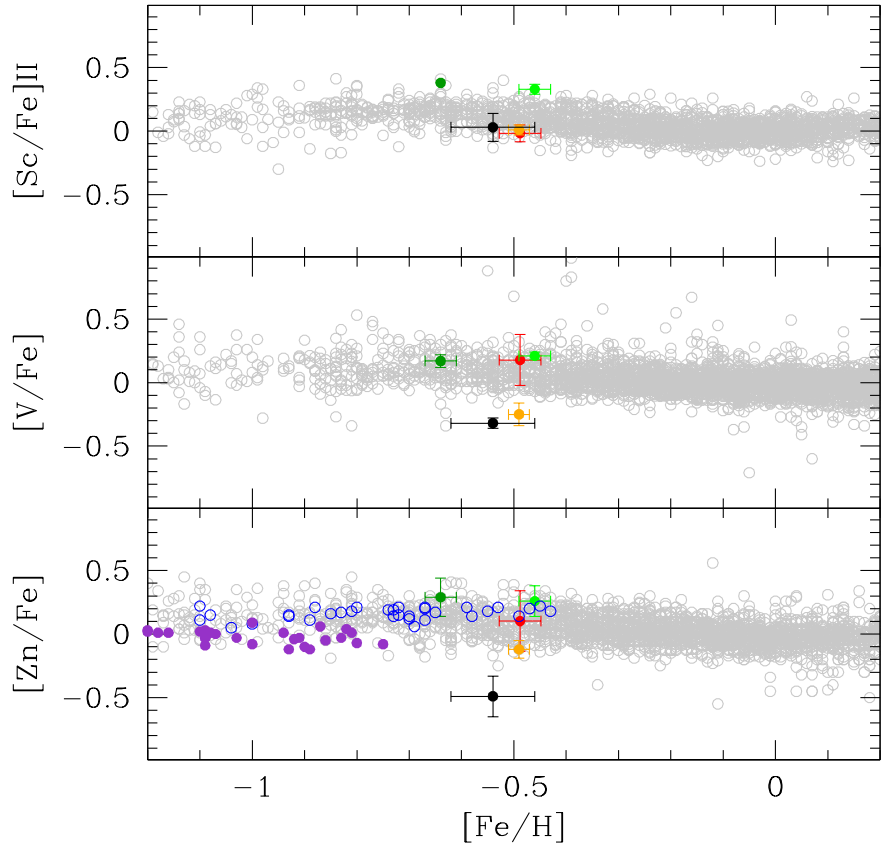}
\caption{Mean abundance ratios and rms scatters for [Sc/Fe], [V/Fe], and 
[Zn/Fe] in NGC~6388 from the present study (red points with errorbars) and in NGC~6388 (orange
points), NGC~6441 (black points), NGC~5927 (light green points), and NGC~6496
(dark green points) from Minelli et al. (2021b). Empty grey circles are field  disc
and bulge stars from  Adibekyan et al. (2012), Battistini and Bensby (2015),
Bensby et al. (2005, 2014, 2017), Bihain et al. (2004), Brewer et al. (2016), da
Silveira et al. (2018), Delgado-Mena et al. (2017), Gratton et al. (2003),
Ishigaki et al. (2012, 2013), Lomaeva et al. (2019), Lucey et al. (2019), and
Reddy et al. (2003, 2006). Open blue circles and filled violet circles are the
high-$\alpha$ and low-$\alpha$ stars in the local sample by Nissen and Schuster
(2011).}
\label{f:fig1}
\end{SCfigure*}

\begin{figure*}
\centering
\includegraphics[scale=0.90]{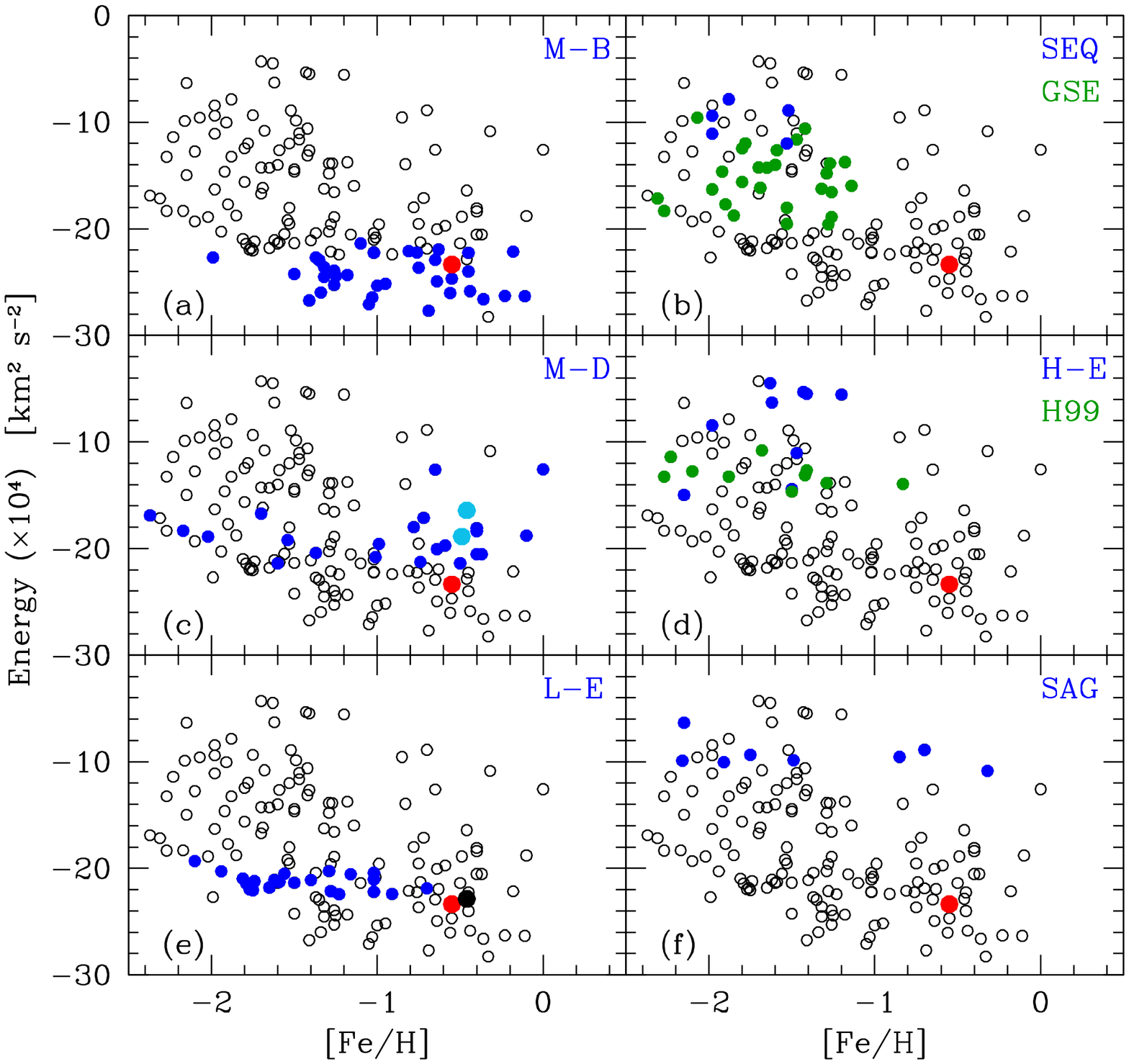}
\caption{Orbital energy E as a function of the metallicity [Fe/H]  for 147 GCs
(empty circles) with orbital parameters from Savino and Posti (2019) and
a kinematical classification in Massari et al. (2019). The different panels show
the position of NGC~6388 (red filled circle), NGC~6441 (black filled circle,
panel (e)), and NGC~5927 and NGC~6496 (cyan circles, panel (c)). In each panel,
GCs associated to different merger events or those that formed in situ  (Massari et al.
2019) are also indicated. The acronyms have the following meanings: main bulge (M-B), main disc
(M-D), low energy (L-E), Sequoia (SEQ), Gaia-Sausage-Enceladus (GSE), high
energy (H-E), Helmi streams (H99), and Sagittarius (SAG).
}
\label{f:fig2}
\end{figure*}

\section{The data set}

The data are those from our project on NGC~6388 `re-loaded', where we exploited
the richness of spectra in the ESO archive to assemble a data set of 
185 highly probable
members in this GC, heavily contaminated by bulge stars. They were analysed
exactly as in our FLAMES survey (see e.g. Carretta et al. 2006 and
Carretta et al. 2010a) and are based on high resolution UVES spectra or GIRAFFE
HR13 spectra. Line lists, solar reference abundances, and corrections for
a hyperfine structure for Sc and V are from Gratton et al. (2003).

Atmospheric parameters and metallicities for the whole sample were
presented in Carretta and Bragaglia (2021a). Individual abundances of Sc, V, and
Zn from archival UVES spectra were partly shown in Carretta and Bragaglia 
(2018). These data are complemented by abundances for individual stars from
our newly acquired UVES (12 stars) and GIRAFFE HR13 spectra 
(150 stars). All the data not shown in Carretta and Bragaglia (2018) will be
presented and discussed in the final paper of this series, mainly devoted to the
properties of multiple stellar populations in NGC~6388.

The abundances were derived using equivalent widths (EWs). Those measured 
in
GIRAFFE spectra were shifted to the system of UVES EWs using stars observed
with both spectrographs (Carretta and Bragaglia, in prep.).
In total, our samples for the present work include 185 stars with a Sc abundance
from UVES and GIRAFFE spectra, 35 stars with a V abundance, and 31 stars with a Zn
abundance. The resulting mean abundance ratios are [Sc/Fe]$=-0.02$ dex
(rms=0.07 dex), [V/Fe]$=+0.26$ dex (rms=0.14 dex), and [Zn/Fe]$=+0.10$ dex
(rms=0.24 dex). 

On average, abundances of Sc were obtained from eight lines for UVES spectra and
two lines for GIRAFFE spectra. Abundances of V and Zn, which are only available for stars
with UVES spectra,  were derived from 12 lines and only one line (Zn~{\sc i}
4810.54~\AA), respectively. Internal errors, estimated with our usual procedure
(see the Appendices in Carretta et al. 2009a,b), are 0.074 dex and 0.088 dex for
Sc (for UVES and GIRAFFE, respectively), 0.074 dex for V, and 0.190 dex for Zn
(only one transition available). There is no trend as a function of the
effective temperature in any of the element abundances.

A detailed comparison of abundances in NGC~6388 is only feasible for Zn, since
we in addition to M21b used the same line. Taking differences
 into account due to the adopted solar reference abundances and the scale of atmospheric
parameters, the final [Zn/Fe] ratios would be virtually the same in the two
analyses (see Appendix A).

The Sc abundances were already used in Carretta and Bragaglia (2019) to
define a robust upper limit to the inner temperature reached by the putative
polluters of the first generation (FG) stars in NGC~6388, that is those that likely
enriched the proto-cluster environment in products of the proton-capture
reactions in H-burning at a very high temperature (see the review by Gratton et
al. 2019).
When we compared the abundances of Mg and Sc for 185 stars in NGC~6388 to the
pattern of field stars from Gratton et al. (2003), we found that the
distribution of cluster stars was an almost perfect match to the field stars
(Carretta and Bragaglia 2019). 
While providing strong constraints on the physical properties of FG polluters in
NGC~6388, the above comparison implicitly showed that no significant difference was
found between the cluster and field stars for one of the three
species claimed by M21a to be a good indicator of nucleosynthesis associated to
extragalactic and/or accreted objects. 

\section{Results}
 
In Carretta and Bragaglia (2019) we only compared the Sc pattern with the sample
of field stars by Gratton et al. (2003). To better ascertain the membership of
NGC~6388 to the autochthonous stellar populations of the MW, in Fig.~\ref{f:fig1} (upper panel), we compare the average [Sc/Fe] abundance measured from
185 cluster stars to different abundance analyses of field stars both in
the Galactic disc (the majority of samples) and the Galactic bulge, in the
metallicity range from [Fe/H]=$-1.2$ dex to 0.2 dex, centred on the mean metal
abundance of NGC~6388. As also stated by M21b, the abundance ratios of iron-peak
elements of disc and bulge stars are nearly identical (see also Griffith et al.
2021). The comparison is extended to V and Zn abundances in the
middle and lower panels, respectively. In each panel, the mean values and
rms scatters from M21b are also indicated for reference.

Considering the intrinsic dispersion associated to the mean values, there is no
evidence of a significant difference between NGC~6388 (red and orange circles 
in Fig.~\ref{f:fig1} correspond to our study and that of  M21b, respectively) and field stars for Sc. Even
NGC~6441 (M21b, black circle) agrees with the pattern of field stars; whereas,
for NGC~5927 and NGC~6496 (the GCs of in situ origin), the Sc seems to be overabundant with respect to the field stars, although still roughly compatible with
the field distribution.

For V, our value based on 35 stars lies inside the field star distribution at
the same level of the mean values for NGC~5927 and NGC~6496. NGC~6441 (two
stars) and NGC~6388 (four stars) from M21b lie below the field star
distribution.

Concerning the Zn abundance, all the mean values for the GCs are in good
agreement with the pattern defined by field disc and bulge stars of the Galaxy,
except for NGC~6441 from M21b.  We would like to caution readers that the results for
NGC~6441 -- in particular those for Zn, resting on a single line lying very
close to the blue spectrum border of the lower signal-to-noise ratio (S/N) -- could be affected by a S/N lower
than optimal, as only about a third of the requested observations were actually
obtained (see Carretta and Bragaglia 2021b).

In summary, our [Sc/Fe] ratio lies at the lower end of the
distribution of field stars with a metallicity similar to that of NGC~6388, while
the V abundance is close to the upper end. However, both values are compatible with
the field stars' distribution; in terms of the standard deviation, neither exceeds
1.7$\sigma$ from the disc/bulge mean value centred on the metallicity of
NGC~6388. On the other hand, for Zn, which is claimed to be the most sensitive indicator
of the accretion's chemical pattern (M21a,b), our average [Zn/Fe] ratio is in perfect
agreement with the bulk of stars in the Galaxy.

Moreover, in many analyses of disc/bulge stars that we used as a 
comparison, no
star is present in the low-Zn region around [Zn/Fe]$\sim -0.5$ dex. Were an
accreted galaxy to produce a set of GCs with such a low Zn level, we  would
also expect the presence of debris consisting of accreted field stars with a
similarly low Zn pattern, which are not observed here. Stars in the MW 
actually do reach such a low Zn content, but they are  bona fide bulge stars,
which are exclusively confined at a metallicity around and above [Fe/H]$=0.0$ dex (da
Silveira et al. 2018), and more than 0.5 dex higher than the metallicity of 
NGC~6388 or the Sag and LMC stars observed in M21a. The low$-\alpha$ stars
in the solar vicinity by Nissen and Schuster (2011) reach a plateau of
[Zn/Fe]$\sim -0.030$ dex (rms=0.057 dex) above [Fe/H]$>-1.1$ dex (see
Fig.~\ref{f:fig1}).  However, the absence of evidence is not evidence of 
an absence,
and perhaps we are seeing a void  simply due to uncomplete or unfortunate
sampling of Galactic populations. In any case, the results of our analysis seem
to exclude an accreted origin  for NGC~6388.

\begin{figure}
\centering
\includegraphics[scale=0.40]{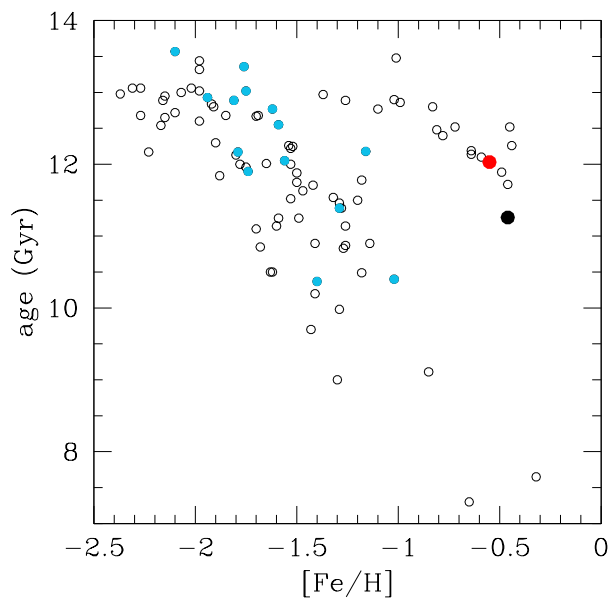}
\caption{Age-metallicity relation for a sample of GCs with an age from Kruijssen et
al. (2019). We highlight  the L-E GCs in Massari et al. (2019) as filled light blue points, NGC~6388 (red point), and NGC~6441 (black point).}
\label{f:fig3}
\end{figure}

\begin{SCfigure*}
\centering
\includegraphics[scale=0.60]{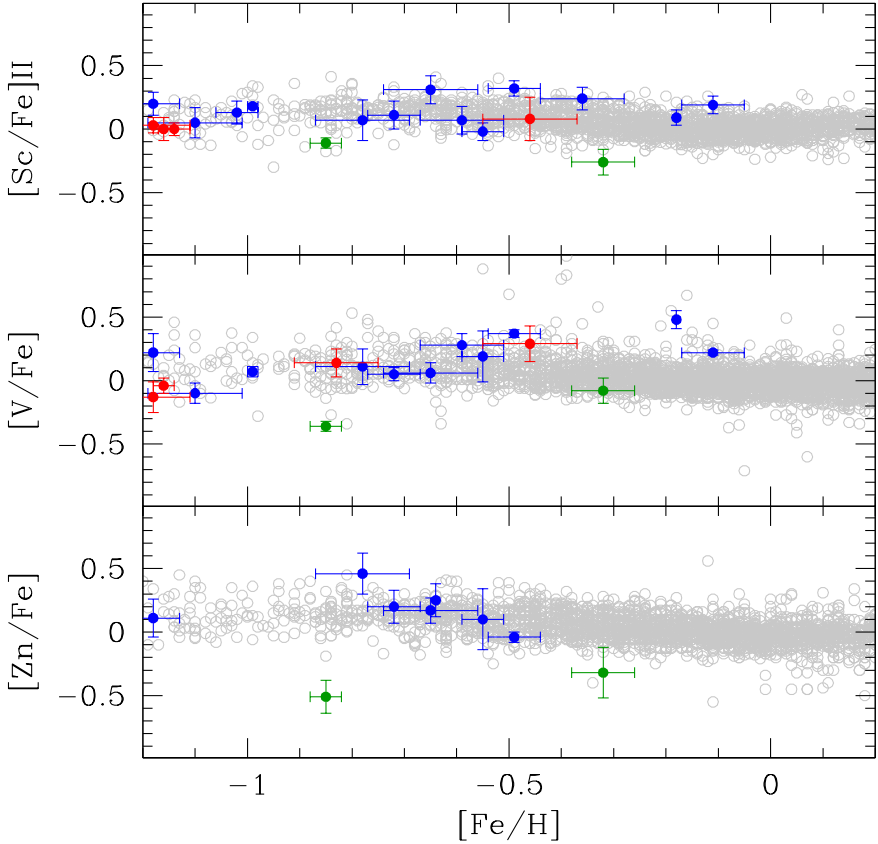}
\caption{Mean abundance ratios [Sc/Fe], [V/Fe], and [Zn/Fe] for metal-rich GCs
in the literature (blue filled circles are for GCs of the main progenitor, either M-B or
M-D; green circles are for Sagittarius GCs, and red circles are for all other accreted
GCs of the L-E, H-E, H99, GSE, and SEQ groups) compared to the field star
distributions. Abundances for GCs are from Carretta (2015), Carretta et al.
(2004, 2011), Cohen (2004), Crestani et al. (2019), Feltzing et al. (2009),
Gratton et al. (2006, 2007), Ivans et al. (1999), Massari et al. (2017), Monaco
et al. (2018), Mu\~{n}oz et al. (2017, 2018, 2020), Mura-Guzm\`an et al. (2018),
O'Connell et al. (2011), Puls et al. (2018),  Ram\`irez and Cohen (2002), 
Sakari et al. (2011),  Sbordone et al. (2007), and Yong et al. (2014).}
\label{f:fig4}
\end{SCfigure*}

\section{Discussion and conclusions}

Our abundance analysis of Sc, V, and Zn abundances for a relevant number of
stars in NGC~6388 shows a chemical pattern virtually indistinguishable from
that of field disc and bulge stars of our Galaxy at a similar metallicity. We
anticipate that also the average [Si/Fe] ($\sim 0.30$ dex from 184
stars, Carretta and Bragaglia, in prep.) supports the in situ formation of
NGC~6388, which is in disagreement with the conclusion by Horta et al. (2020), who assign
it to Sequoia. Their association was made on the basis of the Si 
abundance of a smaller number of stars (24, but with only six
meeting the quality criteria defined by M\'esz\'aros et al. 2020) in the 
infrared data by APOGEE.

To help discriminate between different conclusions about the ancestral origin
 of NGC~6388, we used the chemo-dynamical plane shown
in Fig.~\ref{f:fig2}. We plotted the orbital energy E from Savino and Posti (2019;
we verified that the results did not change using the set by Massari et al. 2019)
as a function of the metallicity of the GCs from Harris (1996, using the 2010 online
edition, which adopts the metallicity scale defined in Carretta et al. 2009c).
Since the integrals of motion such as the orbital energy remain constant along an
orbit, the clumping of GCs around a given energy level may indicate that they
share(d) common orbital paths and likely had the same origin in a common
ancestral system. This is basically the observation used by Massari et al.
(2019), Forbes (2020), Myeong et al. (2019), among others to pick up
coherent ensembles of GCs associated to past merger events. To these IOMs, we added
the chemical dimension represented by the metallicity to help discern
between different origins (similar plots were used in the past, but only to
characterise metal-poor and metal-rich GCs, see for instance Posti and Helmi
2019; Woody and  Schlaufman 2021).
In each panel in Fig.~\ref{f:fig2}, GCs with a common origin according to
Massari et al. (2019) are marked as filled circles, corresponding to the group
labelled inside the panel (see the figure caption for identifications).

The broad anti-correlation depicted in Fig.~\ref{f:fig2}, with high energy
GCs being less metal-rich, on average, can be explained by the tendency of more
highly bound GCs being found in the inner central regions of the Galaxy, in
particular in the bulge where more metal-rich GCs are preferentially located.
GCs with lower binding energies are preferentially confined to lower
metallicities.

Using Fig.~\ref{f:fig2} together with the guidelines from the dynamical analysis by
Massari et al. (2019), it is easy to recognise that NGC~6388 is more likely 
compatible with the main bulge component defined by Massari et al.
Its location in this plane also makes NGC~6388 incompatible with any of the
accretion events known so far, also including those without a well-identified
progenitor, such as the high energy (H-E) GCs. 

In particular, NGC~6388 (as well as NGC~6441) seems to be barely compatible
with the group of GCs with a low orbital energy which have an unknown origin in
Massari et al. (2019) and that are tentatively identified with Kraken (Kruijssen et
al. 2019) or Koala (Forbes 2020). This is hardly surprising since the L-E group
has orbital energies very similar to those of main progenitor GCs, so that L-E
GCs can be recognised as accreted objects only because they neatly lie on the
accretion branch in the bifurcated age-metallicity relation (AMR; e.g. Forbes
and Bridges 2010, Leaman et al. 2013) of MW GCs. A large dispersion in
metallicity (shown in Fig.~\ref{f:fig2}) and a high age normalisation (shown
in Massari et al. 2019) would suggest the existence of a putative massive
candidate progenitor whose debris are still unidentified so far. 
We, however, disagree with the association of NGC~6388 to the L-E group because 
the location of the cluster  is only marginally compatible with them, whereas it
is indistinguishable from other GCs that formed in situ in the bulge of
the main progenitor (panel (a) of Fig.~\ref{f:fig2}).
Hence we confirm the attribution of Massari et al. (2019) and Forbes (2020).

To support the view of NGC~6388 as an autochthonous GC born in our Galaxy and
to exclude the tentative (yet more marginal) association we found with
the L-E group definitively, we show in Fig.~\ref{f:fig3} the AMR of MW GCs, which is an additional
diagnostic plot required to solve the degeneracy concerning the origin of
NGC~6388. We used cluster ages for 96 GCs, with orbital parameters in Savino and
Posti (2019), assembled and homogenised by Kruijssen et al. (2019). We highlight
the GCs of the L-E group, NGC~6388, and NGC~6441. Massari et al. (2019) and
Forbes (2020) used a simple leaky box chemical evolution model to successfully
reproduce the AMR of groups of GCs with a common origin. However, the location of
NGC~6388 and NGC~6441 in Fig.~\ref{f:fig3} implies that a unique simple model
cannot pass through these two GCs and simultaneously through the group of L-E
GCs.

By coupling Fig.~\ref{f:fig2} and Fig.~\ref{f:fig3}, we conclude that the
apparent closeness of NGC~6388 to the L-E GCs in the chemo-dynamical plane E versus
[Fe/H] is misleading. These plots not only confirm the assignment of NGC~6388 to
the M-B group, but they may indicate that NGC~6441 could have also been erroneously
assigned to the L-E group, within which it seems to occupy a similarly marginal
position in panel (e) in Fig.~\ref{f:fig2}.

In Appendix B, we exploit this set of diagnostic plots to discuss some
extant uncertainties in the assignment of GCs to possible past accretion events.
We start from Massari et al. (2019) and Forbes (2020) and are able to solve at
least some of the discrepancies; of course, this will need to be revisited when
more advanced releases of Gaia data are available.

Concerning the chemistry of iron-peak elements as a new criterium for
selecting possible accreted GCs, we perused the literature for abundances of Sc,
V, and Zn in metal-rich GCs. Mean values are plotted in the three panels of
Fig.~\ref{f:fig4}. The only GCs with abundance ratios of these species evidently
below the distribution of Galactic stars are \object{Pal 12} (at [Fe/H]$=-0.82$
dex) and \object{Terzan 7} (at [Fe/H]$=-0.32$ dex; green points), which are both
associated to Sag. The set of chemical tools (Sc, V,and Zn) individuated by
M21a,b seems to be rather efficient in selecting Sag GCs, due to the peculiar
low abundances of these elements in this dwarf galaxy (e.g. Sbordone et al.
2007). 

The use of iron-peak elements for picking up accreted GCs 
seems to be less evident, in general, and not yet fully proven. The GCs assigned to other 
accreted components (red points in Fig.~\ref{f:fig4}) do not seem to have lower
abundances than MW stars.  Moreover, when examining mean values of
[Sc/Fe] (the most commonly studied element among Sc, V, and Zn in GCs) collected
for a large sample of  GCs of any metal abundance in Carretta and Bragaglia
(2021b), we again found that the average abundances of accreted GCs -- apart from a
couple of exceptions -- are a perfect match to the MW pattern down to
[Fe/H]$\sim -2.5$ dex.

The efficiency of the method proposed by M21a,b should be
tested on a larger sample of GCs, to be analysed homogeneously to the comparison
sample. This is something the large spectroscopic surveys in the optical range,
such as WEAVE (Dalton et al. 2020) and 4MOST (de Jong et al. 2019), will
efficiently provide in the near future.\\

\begin{acknowledgements}
We are grateful to Alessandro Savino and Davide Massari for sharing their tables
of orbital parameters. 
This research has made use of the services of the ESO Science Archive Facility,
of the SIMBAD database (Wenger et al. 2000), operated at CDS, Strasbourg,
France, and in particular of the VizieR catalogue access tool (Ochsenbein et al.
2000). This research has made use of NASA's Astrophysical Data System.
\end{acknowledgements}

\begin{appendix}

\section{Comparison with M21b for Zn}

The four stars analysed in NGC~6388 by M21b were also studied in Carretta et al.
(2007) and homogeneously re-analysed in Carretta and Bragaglia (2018). Here we
show a detailed comparison of the Zn abundances in M21b; since they are only based on one line, Zn~{\sc i} 4810.54~\AA\, is used in both studies.

In Table~\ref{t:tabapp} we list the atmospheric parameters and abundances from M21b
and from our analysis. The different sets of adopted solar reference abundances
imply that offsets of $-$0.04 dex and +0.05 dex for Fe and Zn,  respectively, 
should be applied to abundances in M21b to bring their values onto our scale.
In addition, an offset of +0.08 dex and +0.07 dex for Sc and V, respectively, 
would bring the abundances of Sc and V in M21b onto our scale, reducing
the difference from field stars.

Another relevant source of differences between ours and those of M21b is the scale of
atmospheric parameters used in the abundance analysis. M21b used a
semi-spectroscopic set, while our method is entirely based on a photometric
approach (see Carretta and Bragaglia 2018, 2021a). As an example, for the four
stars under scrutiny, the effective temperatures in M21b are higher on average
by $\sim 88$ K, with an rms scatter of 11 K.

In Table~\ref{t:tabapp2} we report the sensitivities of Fe and Zn abundances to
variations in the atmospheric parameters from Carretta and Bragaglia (2018). We
used these sensitivities, together with the observed differences in the
atmospheric parameters from Table~\ref{t:tabapp}, to compute the average offset
required to bring the [Zn/Fe] abundances by M21b onto our scale of atmospheric
parameters. The offset turns out to be +0.043 dex.

Finally, also considering the difference in the adopted oscillator strength  for
the Zn line ($-0.15$ in M21b from Roederer and Lawler 2012; and $-0.17$ in our
analysis, from Bi\'emont and Godefroid 1980), the Zn average abundance by M21b
would be on our scale [Zn/Fe]$=-0.12+0.05+0.043+0.02 = -0.008$ dex, which is in good
agreement with our mean value $-0.02$ dex.

\begin{table}[h]
\small
\centering
\setlength{\tabcolsep}{0.5mm}
\caption[]{Comparison of our study with M21b for four stars in NGC~6388.}
\begin{tabular}{lcccccccccc}
\hline
star   & T$_{\rm eff}$ & $\log g$ & [Fe/H]   & $v_t$  &  [Zn/Fe] & T$_{\rm eff}$ & $\log g$ & [Fe/H]   & $v_t$ &  [Zn/Fe] \\
 &\multicolumn{5}{c}{M21b} &\multicolumn{5}{c}{us} \\
\hline        
u63a   &          4100 &     1.33 &  $-$0.49 &  1.60  &  $-$0.07 &  4018          & 1.37     & $-$0.407 & 1.66  &   +0.323 \\
u63b   &          4150 &     1.42 &  $-$0.46 &  1.50  &  $-$0.21 &  4046          & 1.42     & $-$0.415 & 1.75  & $-$0.066 \\
u63e   &          4000 &     1.16 &  $-$0.51 &  1.50  &  $-$0.12 &  3913          & 1.15     & $-$0.457 & 1.65  &   +0.173 \\
u63f   &          4000 &     1.16 &  $-$0.50 &  1.50  &  $-$0.07 &  3922          & 1.17     & $-$0.414 & 1.57  &   +0.041 \\
                                                                           
\hline
\end{tabular}
\begin{list}{}{}
\item[Solar reference abundances:]
M21b: Fe 7.50 and Zn 4.60 are from Grevesse and Sauval (1998). For our study: Fe 7.54 and Zn 4.59 are from Gratton et al. (2003)
\end{list}
\label{t:tabapp}
\end{table}

\begin{table}[h!]
\centering
\caption[]{Sensitivities of abundance ratios to variations in the atmospheric
parameters for Fe and Zn.} 
\begin{tabular}{lrrrr}
\hline
Element                 & T$_{\rm eff}$ & $\log g$ & [A/H]   & $v_t$      \\
                        &          (K)  &  (dex)   & (dex)   & kms$^{-1}$ \\
\hline        
Variation               &  50           &   0.20   &  0.10   &  0.10      \\
\hline
$[$Fe/H$]${\sc  i}      & $-$0.006      & +0.040   & +0.024  & $-$0.045   \\
$[$Zn/Fe$]${\sc i}      & $-$0.033      & +0.011   &  0.000  & $-$0.011   \\
                                                                           
\hline
\end{tabular}
\label{t:tabapp2}
\end{table}

\section{GCs with an uncertain attribution in Massari et al. (2019)}

There are a few uncertainties in the association of GCs to coherent groups or
components sharing the same origin in Massari et al. (2019, Ma19 in this
appendix), that is the paper we used as a reference here. The  attribution of some
GCs was considered unsure and indicated by a question mark or there were multiple options 
in their Table~1. Some of these cases were discussed by Forbes (2020, F20 in
this appendix), for whom the more extended set of ages in Kruijssen et
al. (2019, K19 in this appendix) were available, the same we are using here. It is possible that these
uncertainties will be resolved using improved kinematics based on future Gaia
data releases, but in the meantime we tried to alleviate the involved degeneracy
employing the chemodynamical plots E versus [Fe/H] and L$_Z$ versus [Fe/H], together
with the AMR (whenever accurate ages are available). 

\paragraph{\object{Pal~2}:} The cluster is tagged as 'GSE?' in Ma19 and also F20
has the same indication. This GC is surely an accreted object, compatible with
Seq, H99, and GSE (see Fig.~B.1, left column); no age is available in K19; the
Lz-Fe plot shows it to be  more probably of GSE origin, which is 
compatible with H99.

\paragraph{\object{NGC 3201}:} The cluster is marked as 'Seq/GSE' in Ma19, while F20 puts
it in the Seq sample; the uncertainty is also confirmed by the two upper plots
(Fig.~B.1, right column); however, the Lz-Fe plot would indicate a more probable
Seq origin. Gaia DR3 and later releases are required to settle this issue.

\paragraph{\object{E 3}:} The cluster is indicated by 'H99?' in Ma19, while F20
puts it among the in situ GCs. Looking at the upper left panel of Fig.~B.2, H99
is a possibility, but also M-D; the AMR indicates a clear in situ origin and the
Lz-Fe plot is in agreement, even if more marginally so. We thus agree with F20. 

\paragraph{\object{NGC 5139} ($\omega$~Cen:)} The cluster is 'GSE/Seq' in Ma19
and Seq in F20; our plots (Fig.~B.2, right column) indicate a strong preference
for a GSE origin.

\paragraph{\object{Rup 106}:} The cluster is marked as 'H99?' in Ma19 and H99 in
F20; the attribution seems to be confirmed by the plots in Fig.~B.3 (left
column).

\paragraph{\object{Pal 5}:} The cluster is 'H99?' in Ma19 and H99 in F20;  the
attribution seems clear from the Lz-Fe plot in Fig.~B.3 (right column).

\paragraph{\object{NGC 5634}:} The cluster is 'H99/GSE' in Ma19 and H99 in F20;
looking at the three plots in Fig.~B.4 (left column), the ambiguity in
attribution still seems unresolved.

\paragraph{\object{NGC 5904}:} The cluster is 'H99/GSE' in Ma19 and H99 in F20;
also in this case, the three plots in Fig.~B.4 (right column) are not resolutive,
although a slight preference for GSE seems possible from the Lz-Fe plane. 

\paragraph{\object{NGC 6101}:} The cluster is 'Seq/GSE' in Ma19 and Seq in F20;
while both can be possible, as can be seen from Fig.~B.5 (left panels), a rather strong 
preference for Seq is visible in the bottom plot.     

\paragraph{\object{Pal 15}:} The cluster is 'GSE?' in Ma19 and GSE in F20; the
two upper panels of Fig.~B.5 would also allow a Sag origin, while the lower
panel seems to suggest that GSE is favoured.

\paragraph{\object{NGC 6535}:} The cluster is 'L-E/Seq' in Ma19 and Seq in F20;
the three plots on the left in Fig.~B.6 leave no doubts as to a L-E
origin for this GC.

\paragraph{\object{Liller 1}:} The cluster was not attributed to any group
in Ma19, while F20 gives it an in situ origin. We agree with the latter
(see Fig.~B.6, right panels), with a M-B origin for this GC.

\paragraph{\object{NGC 6426}:} The cluster was attributed to the H-E group both
by Ma19 and F20; however, the plots in Fig.~B.7 seem to put this into question, 
favouring a possible H99 origin.

\begin{figure*}
\centering
\includegraphics[scale=0.55]{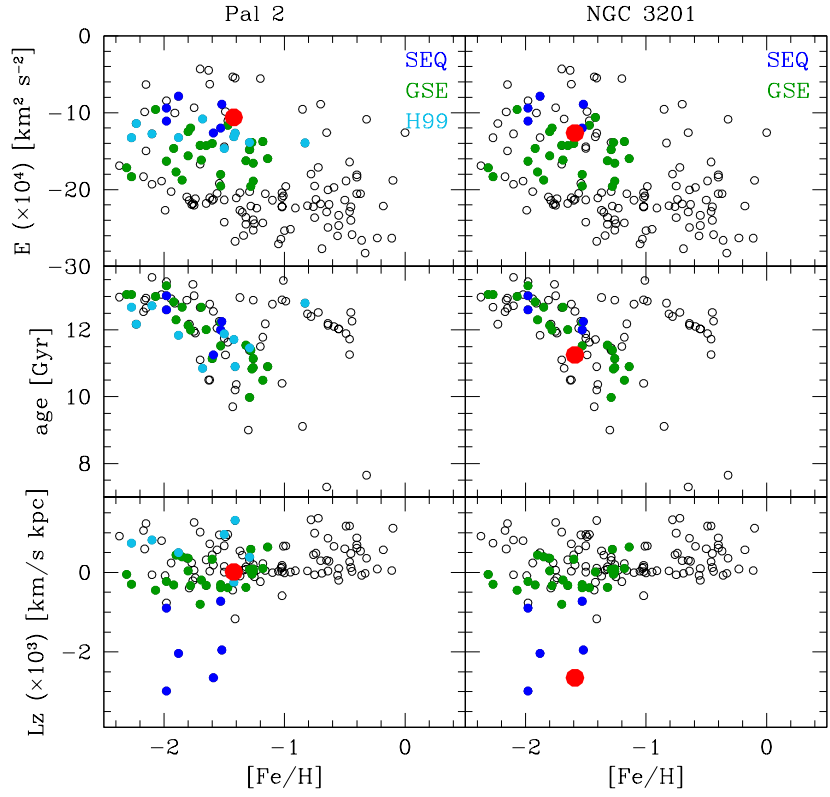} 
\caption{Diagnostic chemo-dynamical plots (upper and lower panels) and AMR 
(middle panel) for two GCs (Pal~2 and NGC~3201) with an uncertain attribution to different groups in Ma19 and F20.}
\label{f:appfig1}
\end{figure*}

\begin{figure*}
\centering
\includegraphics[scale=0.55]{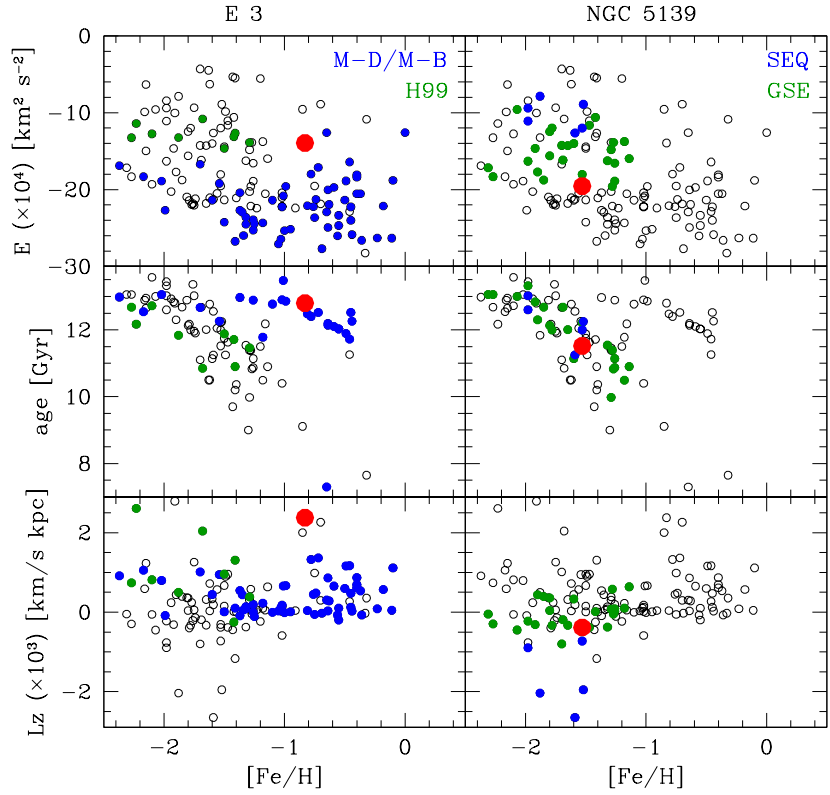} 
\caption{Same as Fig.\ B.1, but for E3 and NGC~5139 ($\omega$ Cen).}
\label{f:appfig2}
\end{figure*}

\begin{figure*}
\centering
\includegraphics[scale=0.55]{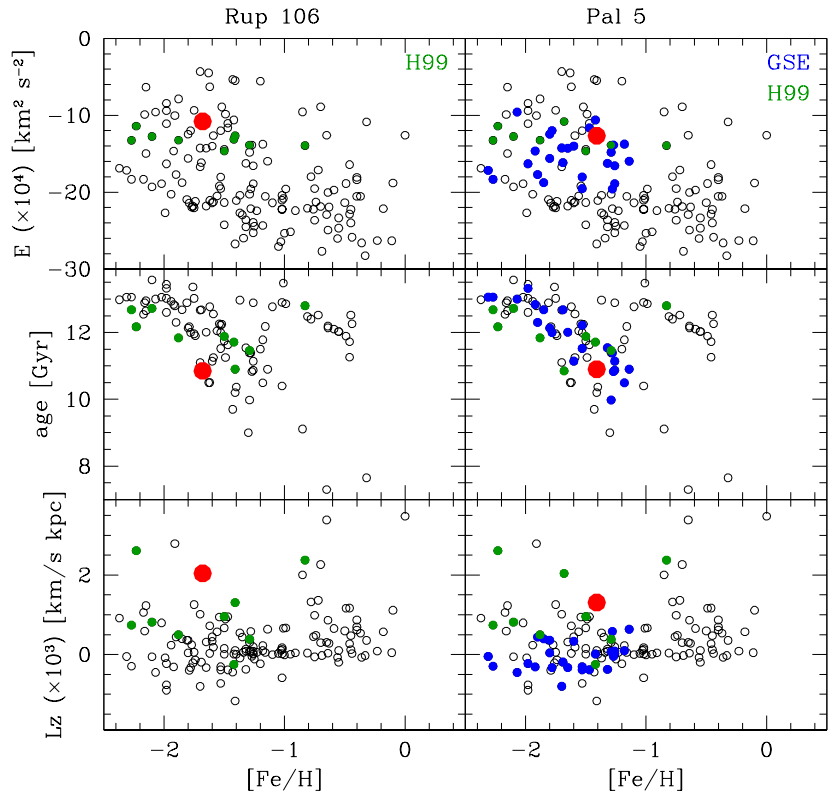} 
\caption{Same as Fig.\ B.1, but for Rup~106 and Pal 5.}
\label{f:appfig3}
\end{figure*}

\begin{figure*}
\centering
\includegraphics[scale=0.55]{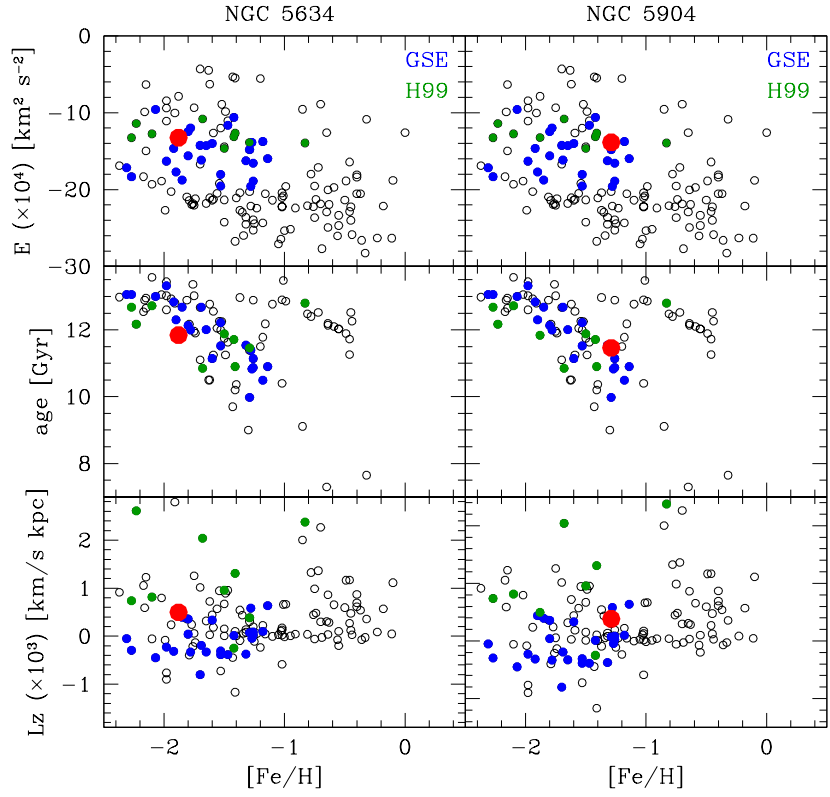} 
\caption{Same as Fig.\ B.1, but for NGC~5634 and NGC~5904.}
\label{f:appfig4}
\end{figure*}

\begin{figure*}
\centering
\includegraphics[scale=0.55]{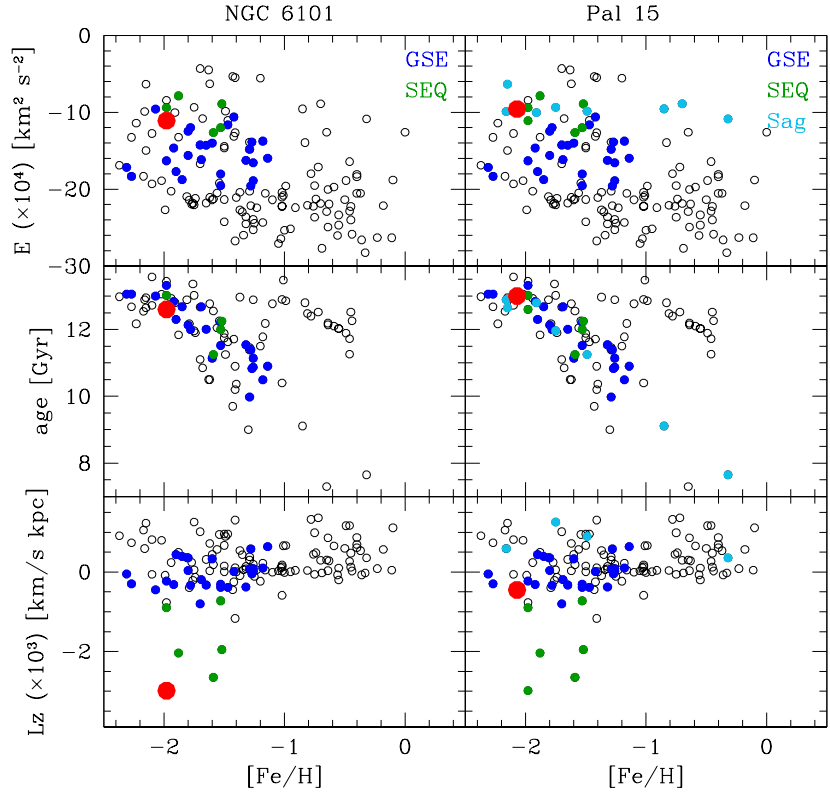} 
\caption{Same as Fig.\ B.1, but for NGC~6101 and Pal 15.}
\label{f:appfig5}
\end{figure*}

\begin{figure*}
\centering
\includegraphics[scale=0.55]{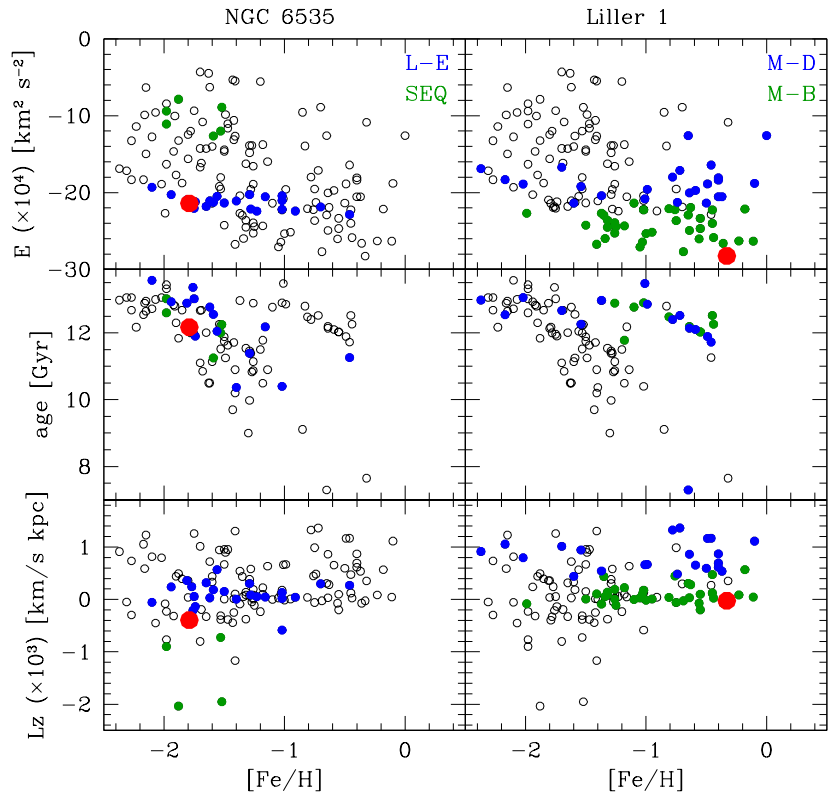} 
\caption{Same as Fig.\ B.1, but for NGC~6535 and Liller 1.}
\label{f:appfig6}
\end{figure*}

\begin{figure*}
\centering
\includegraphics[scale=0.55]{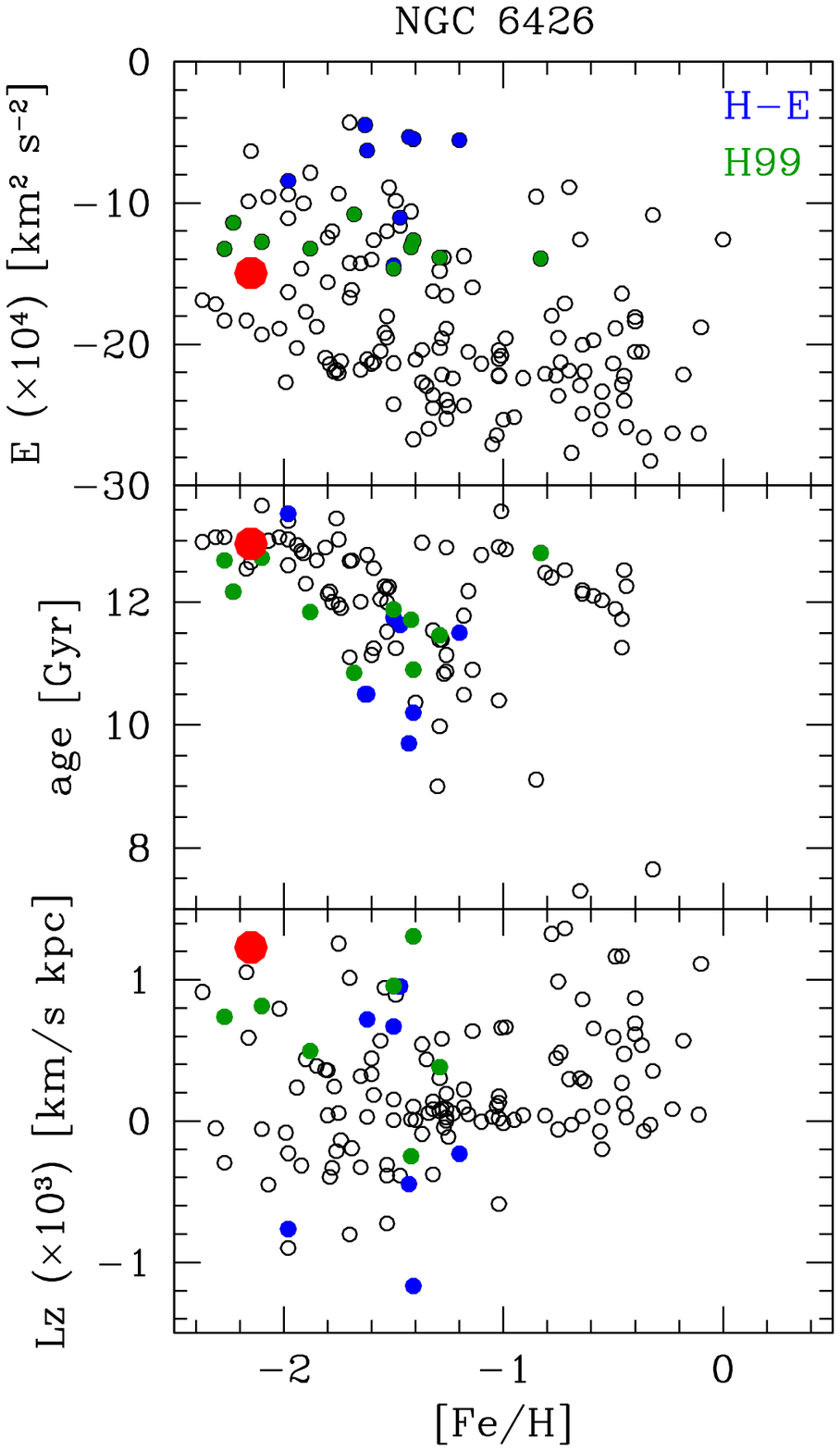} 
\caption{Same as Fig.\ B.1, but for NGC~6426.}
\label{f:appfig7}
\end{figure*}

\end{appendix}


\begin{thebibliography}{}

 \bibitem[]{} Adibekyan, V.Zh., Sousa, S.G., Santos, N.C. et al. 2012, A\&A, 545,
  A32
\bibitem[]{} Battistini, C., Bensby, T. 2015, A\&A, 577, A9
\bibitem[]{} Bensby, T., Feltzing, S., Lundstr{\"o}m, I., Ilyin, I. 2005, A\&A,
  433, 185
\bibitem[]{} Bensby, T., Feltzing, S., Oey, M.S. 2014, A\&A, 562, A71
\bibitem[]{} Bensby, T., Feltzing, S., Gould, A. et al. 2017, A\&A, 605, A89
\bibitem[]{} Bi\'emont, E., Godefroid, M. 1980, A\&A, 84, 361 
\bibitem[]{} Bihain, G., Israelian, G., Rebolo, R., Bonifacio, P., Molaro, P.
  2004, A\&A, A423, 777
\bibitem[]{} Brewer, J.M., Fischer, D.A., Valenti, J.A., Piskunov, N. 2016,
  ApJS, 225, 32
\bibitem[]{} Carretta, E. 2015, ApJ, 810, 148 
\bibitem[]{} Carretta, E., Bragaglia, A. 2018, A614, A109 
\bibitem[]{} Carretta, E., Bragaglia, A. 2019, A\&A, 627, L7 
\bibitem[]{} Carretta, E., Bragaglia, A. 2021a, arXiv:2111.12721
\bibitem[]{} Carretta, E., Bragaglia, A. 2021b, A\&A, 646, A9 
\bibitem[]{} Carretta, E., Gratton, R.G., Bragaglia, A., Bonifacio, P., 
  Pasquini, L. 2004, A\&A, 416, 925 
\bibitem[]{} Carretta, E., Bragaglia, A., Gratton, R.G. et al. 2006, A\&A, 450, 523 
\bibitem[]{} Carretta, E., Bragaglia, A., Gratton, R.G., Lucatello, S. 2009a, 
 A\&A, 505, 139 
\bibitem[]{}  Carretta, E., Bragaglia, A., Gratton, R.G. et al. 2009b, 
  A\&A, 505, 117  
\bibitem[]{} Carretta, E., Bragaglia, A., Gratton, R.G., D'Orazi, V., Lucatello,
 S. 2009c, A\&A, 508, 695 
\bibitem[]{} Carretta, E., Bragaglia, A., Gratton, R.G. et al. 2010a, A\&A, 516, 55 
\bibitem[]{} Cohen, J.G. 2004, AJ, 127, 1545  
\bibitem[]{} Crestani, J., Alves-Brito, A., Bono, G., Puls, A.A., 
  Alonso-Garc\'ia, J. 2019, MNRAS, 487, 5463 
\bibitem[]{} Dalton, G., Trager, S., Abrams, D.C. et al. 2020, SPIE, 11447, 14 
\bibitem[]{} da Silveira, C.R., Barbuy, B., Fria\c ca, A.C.S. et al. 2018, A\&A,
  614, A149
\bibitem[]{} de Jong, R.S., Agertz, O., Berbel, A.A. et al. 2019, Msngr. 175, 3  
\bibitem[]{} Delgado-Mena, E., Tsantaki, M., Adibekyan, V.Zh. et al. 2017, A\&A,
  606, A94
\bibitem[]{} Feltzing, S., Primas, F., Johnson, R.A. 2009, A\&A, 493, 913 
\bibitem[]{} Feuillet, D.K., Sahlholdt, C.L., Feltzing, S., et al. 2021,
  MNRAS, 508, 1489
\bibitem[]{} Forbes, D.A. 2020, MNRAS, 493, 847
\bibitem[]{} Forbes, J.M., Bridges, T. 2010, MNRAS, 404, 1203 
\bibitem[]{} Freeman, K., Bland-Hawthorn, J. 2002, ARA\&A, 40, 487 
\bibitem[]{} Gratton, R.G., Carretta, E., Claudi, R., Lucatello, S.,
  Barbieri, M. 2003, A\&A, 404, 187 
\bibitem[]{} Gratton, R.G., Lucatello, S., Bragaglia, A. et al. 2007, A\&A, 
  464, 953 
\bibitem[]{} Gratton, R.G., Lucatello, S., Bragaglia, A. et al. 2006, A\&A, 
  455, 271
\bibitem[]{} Gratton, R.G., Bragaglia, A., Carretta, E. et al. 2019, A\&ARv, 27, 8
\bibitem[]{} Grevesse, N., Sauval, A.J. 1998, SSRv, 85, 161
\bibitem[]{} Griffith, E., Weinberg, D.H., Johnson, J.A. et al. 2021, ApJ, 909,
  77
\bibitem[]{} Harris, W.~E. 1996, AJ, 112, 1487
\bibitem[]{} Hasselquist, S., Carlin, J.L., Holtzman, J.A. et al. 2019, ApJ,
  872, 58
\bibitem[]{} Horta, D., Schiavon, R.P., Mackereth, J.T. et al. 2020, MNRAS, 493,
  3363  
\bibitem[]{} Ishigaki, M.N., Chiba, M., Aoki, W. 2012, ApJ, 753, 64
\bibitem[]{} Ishigaki, M.N., Aoki, W., Chiba, M. 2013, ApJ, 771, 67
\bibitem[]{} Ivans, I.I., Sneden, C., Kraft, R.P. et al. 1999, AJ, 118, 1273 
\bibitem[]{} Kruijssen, J.M.D., Pfeffer, J.L., Reina-Campos, M., Crain, R.A., 
  Bastian, N. 2019, MNRAS, 486, 3180 
\bibitem[]{} Leaman, R., VandenBerg, D.A., Mendel, J.T. 2013, MNRAS, 436, 122 
\bibitem[]{} Lomaeva, M., J{\"o}nsson, H., Ryde, N., Schulteis, M., Thorsbro, B.
  2019, A\&A, 625, A141
\bibitem[]{} Lucey, M., Hawkins, K., Ness, M. et al. 2019, MNRAS, 488, 2283
\bibitem[]{} Massari, D., Koppelman, H.H., Halmi, A. 2019, A\&A, 630, L4
\bibitem[]{} Massari, D., Mucciarelli, A., Dalessandro, E. et al. 2017, MNRAS,
  468, 1249 
\bibitem[]{} M\'esz\'aros, S., Masseron, T., Garc\'ia-Hern\'andez, D.A. et al.
  2020, MNRAS, 492, 1641 
\bibitem[]{} Minelli, A., Mucciarelli, A., Romano, D. et al. 2021a, ApJ, 910, 
  114 (M21a)
\bibitem[]{} Minelli, Mucciarelli, A., Massari, D. et al. 2021b, ApJ, 918, 
  L32 (M21b)
\bibitem[]{} Monaco, L., Villanova, S., Carraro, G., Mucciarelli, A., Moni
  Bidin, C. 2018, A\&A, 616, A181 
\bibitem[]{} Mu\~{n}oz, C., Villanova, S., Geisler, D. 2017, A\&A, 605, A12 
\bibitem[]{} Mu\~{n}oz, C., Geisler, D., Villanova, S. et al. 2018, A\&A, 620,
  A96 
\bibitem[]{} Mu\~{n}oz, C., Villanova, S., Geisler, D. et al. 2020, MNRAS, 492,
  3742 
\bibitem[]{} Mura-Guzm\'an, A., Villanova, S., Mu\~{n}oz, C, Tang, B. 2018,
  MNRAS, 474, 4541  
\bibitem[]{} Myeong, G.C., Vasiliev, E., Iorio, G., Evans, N.W., Belokurov, V.
  2019, MNRAS, 488, 1235 
\bibitem[]{} Nissen, P.E., Schuster, W.J. 2010, A\&A, 511, L10
\bibitem[]{} Nissen, P.E., Schuster, W.J. 2011, A\&A, 530, A15 
\bibitem[]{} Ochsenbein, F., Bauer, P., Marcout, J. 2000, A\&AS, 143, 23
\bibitem[]{} O'Connell, J.E., Johnson, C.I., Pilachowski, C.A., Burks, G. 2011,
 PASP, 123, 1139 
\bibitem[]{} Posti, L., Helmi, A. 2019, A\&A, 621, A56
\bibitem[]{} Puls, A.A., Alves-Brito, A., Campos, F., Dias, B., Barbuy, B. 2018,
  MNRAS, 476, 690 
\bibitem[]{} Ram\'irez, S., Cohen, J.G. 2002, AJ, 123, 3277 
\bibitem[]{} Reddy, B.E., Tomkin, J., Lambert, D.L., Allende Prieto, C. 2003,
  MNRAS, 340, 304
\bibitem[]{} Reddy, B.E., Lambert, D.L., Allende Prieto, C  2006, MNRAS, 367,
  1329
\bibitem[]{} Roederer, I.U., Lawler, J.E. 2012, ApJ, 750, 76 
\bibitem[]{} Sakari, C.M., Venn, K.A., Irwin, M. et al. 2011, ApJ, 740, 106 
\bibitem[]{} Sbordone, L., Bonifacio, P., Buonanno, R. et al. 2007, A\&A, 465, 
  815 
\bibitem[]{} Savino, A., Posti, L. 2019, A\&A, 624, L9
\bibitem[]{} Wenger, M., Ochsenbein, F., Egret, D. et al. 2000, A\&AS, 143, 9
\bibitem[]{} Woody, T., and Schlaufman, K.C. 2021, AJ, 162, 42
\bibitem[]{} Yong, D., Alves Brito, A., Da Costa, G.S. et al. 2014, MNRAS, 439,
  2638 
 


\end{thebibliography}
\end{document}